# THE COLORFUL CHEMICAL BOTTLE EXPERIMENT KIT: FROM SCHOOL LABORATORY TO PUBLIC DEMONSTRATION

Taweetham Limpanuparb*, Suphattra Hsu

*Mahidol University International College, Mahidol University, Salaya, Nakhon Pathom, 73170, Thailand*

*E-mail: taweetham.lim@mahidol.edu, Tel. +66 2441 5090 Ext. 3538, Fax. +66 2441 5092

**Abstract:** The blue bottle experiment was first introduced to the chemical education literature as a simple demonstration on kinetics. Its original formulation contains only glucose, NaOH and small amount of methylene blue. The solution turns blue when shaken and fades to colorless upon standing. This bluing/de-bluing cycle may be repeated and may be compared to blood colors in animal's respiratory cycle. Inspired by the blue bottle experiment, the colorful chemical bottle experiment kit was commercially developed in 2006. The kit is a versatile pedagogical tool, not only for physical chemistry but also for analytical, biological and organic chemistry. It also helps teaching concepts in scientific method and laboratory safety. This manuscript contains four parts, brief review on literature relating to the blue bottle experiment, description of the colorful chemical bottle experiment kit, pedagogical discussion of the experiments and preliminary evaluation from students.

## 1. Introduction

The blue bottle experiment was published in the Journal of Chemical Education by Campbell in 1963. He learned of the experiment from the University of Wisconsin but attributed its origin to CalTech. His formulation contains roughly 20 g NaOH, 20 g glucose and half a milliliter of 1% methylene blue (MB) in one liter of solution. The solution change from clear to blue and blue to clear as it is shaken or left to stand. The bluing/de-bluing cycle may be repeated many times [1].

The experiment was introduced to a variety of audiences from 7$^{th}$ grade to graduate students. Without revealing the chemical composition of the solution, Campbell asked them why the solution changes its color. Students proposed hypotheses and methods to verify them. Here are some examples:

• The solution turn blues because it comes into contact with color tainted on the stopper on the top of the flask. This was proven wrong by swirling the flask so that the solution does not touch the stopper.

• The solution turn blues because of frictional heat from shaking the flask. This was proven wrong by warming the flask by hands.

• Shaking the flask combines upper thin blue layer and colorless bottom layer and, as a result, introduces blue color to the whole flask. The two layers form again as the solution is left to settle. This hypothesis contradicts the observation that blue color uniformly disappears in all direction rather than migrating upward.

After a number of trial-and-error practices, students eventually suggested that by shaking the flask blue color is formed by a chemical reaction with the air. This was confirmed by shaking a flask filled with natural gas and the blue bottle solution. The solution did not turn blue. The natural gas was then replaced by atmospheric air and it turned blue again upon shaking.

Inspired by Campbell's work, Engerer and Cook wrote an undergraduate laboratory manual for reaction mechanism consisting of step-by-step series of procedures and questions on the blue bottle experiment. It was hailed favorite experiment of the year during course evaluation at Valparaiso University. Their materials for both students and instructors are available on the journal website [2].

There are a number of modifications to the original formulation (e.g. chemical traffic light, vanishing valentine, variations of sugar and solvent, replacement of strong base and sugar with ascorbic acid for green chemistry), chemical pattern formation and a number of applications of the reaction (e.g. oxygen indicator for intelligent food packaging and oxygen scavenger for super-resolution imaging). We briefly discuss them below.

Methylene blue may be replaced by its derivatives or analogues giving a variety of oxidized colors. A notable example is resazurin in vanishing valentine experiment where the solution is colorless/fluorescent red rather than colorless/blue [3]. Alternatively, a different family of dye can be used. Indigo carmine is used in chemical traffic light reaction that gives yellow, red and green solution [4].

Cook *et al.* reported their modification to the classic blue bottle experiment in his 1994 article "Blue Bottle Experiment Revisited". He experimented with fifteen structurally similar dyes, six reducing sugars and two polar solvents. The varying activation energies observed for the debluing reaction was well explained by anomeric effect [5].

Since the classical blue bottle experiment is rather corrosive and has high sugar content, Wellman proposed an environmentally friendly alternative by using ascorbic acid (vitamin C) and ion catalyst to reduce the dye. The new formulation somewhat preserves color change behavior. It may not be

471

perfect, but improves the overall safety and environmental concerns [6].

At microscopic level, the color change is not homogeneous. When the solution is placed on thin dish, after a period of time, patterns may develop on the surface of the solution [7]. This is due to convection cell formed by different density of glucose and its oxidized product.

Apart from teaching and demonstration tools, we also find recent applications of the blue bottle experiment, for example, oxygen indicator [8] and oxygen scavenger [9].

## 2. Colorful Chemical Bottle Experiment Kit

The colorful chemical bottle experiment kit consists of four chemical bottles (glucose, NaOH, MB and indigo carmine), three blank graduated plastic bottles, four needleless syringes, four plastic stirring rods, label sticker and a booklet (in Thai) in a zip bag. It was developed by Limpanuparb in collaboration with Thailand's National Science and Technology Development Agency (NSTDA). More than 2,000 kits were produced in March 2006 with a recommended retail price of THB99. It was sold and distributed by NSTDA bookstore network. It has been shown to the public at many national and local events including Science Week, Children Day and various educational activities of NSTDA's Sirindhorn Science Home. In 2014, the experiment was also introduced to an integrated laboratory course at the Mahidol University International College (MUIC).

The booklet starts with a flowchart that outlines steps to be taken during the experiment. Before commencing the experiment, the user must read safety precautions and find additional materials for the experiment, PETE bottle, pen, tissue paper, stopwatch, and personal protective equipments (PPEs). Caution should be exercised while handling concentrated NaOH (corrosive) and MB (stainable). Partial Material Safety Data Sheet (MSDS) information is also provided in the booklet. The compulsory materials (PETE bottle and pen) are usually available at home while stopwatch and PPEs may be purchased at grocery/hardware stores.

The first step to the experiment is to prepare four stock solutions by filling water to the mark on provided chemical bottles. Exact weights of chemicals are written on the bottles and concentrations may be calculated according to the third section of the booklet. Provided graduated bottle and label sticker are used to make 50/100ml marks on the PETE bottle.

The second step is to make a conventional blue bottle experiment. The three chemicals are drawn by separate syringes to the PETE bottle. Two spare plastic bottles are provided in the kit to facilitate the transfer of glucose and NaOH. Users will explore the cause of color changes by conducting series of experiments designed by Campbell and Cook.

The next step is to investigate concentration dependence of debluing time. Blank tables and graphs are provided for three sets of experiments to vary the concentration of one chemical at a time. It is designed in way that the time vs concentration plot will be linear.

The final experiment in the kit is to replace MB by indigo carmine. This is only simple qualitative demonstration but the climax of the kit. Users will observe that the blue dye transforms into chemical traffic light of green (oxidized color), red (intermediate) and yellow (reduced color) during the shaking/standing cycle.

For advanced user, supplementary information including redox equations, differential/integrated rate law, Arrhenius equations and references are provided at the end of the booklet. Rate constants and the activation energy may be obtained from the experiments at different temperatures. Though the experiment kit consists of disposable and inexpensive materials, their functions are comparable to analytical glassware like volumetric flask or pipette.

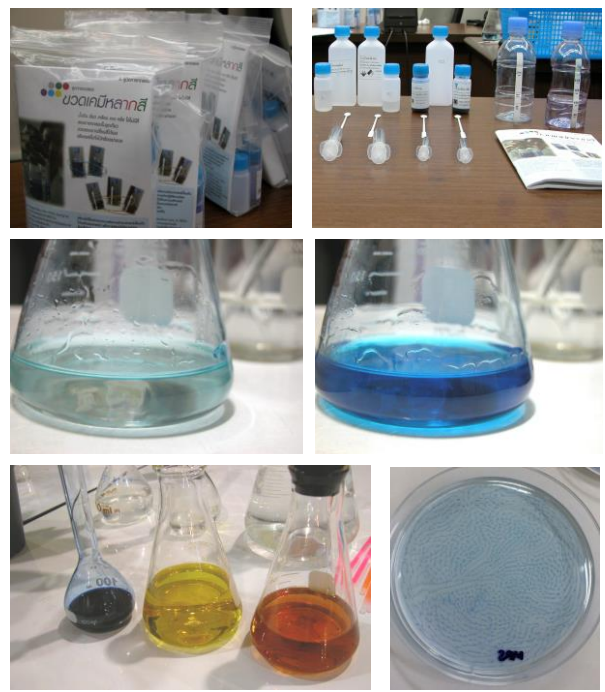

Figure 1. (top) The NSTDA's colorful chemical bottle experiment kit developed by Limpanuparb in 2006, (middle) the classical blue bottle experiment, (bottom-left) the chemical traffic light, and (bottom-right) line and dot patterns formed in the blue bottle experiment.

## 3. Pedagogical functions

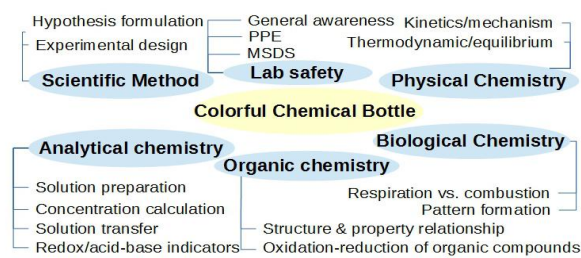

The experiment kit has versatile pedagogical applications. We elaborate possible use of our experiment below.

**3.1 Physical chemistry**

The classical blue bottle experiment is commonly regarded as a demonstration for chemical kinetics. We follow the approach of Campbell and Cook to encourage students to propose reaction mechanisms and verify them through experiments and observations. Since the total reaction of this experiment is only oxidation of glucose by oxygen under basic condition to gluconate ion, the organic dye can be regarded as a catalyst. This is surprising to some students as they may have different preconception of catalysts. However, when it is compared to respiration system (See Biological chemistry section.), most students are able to accept that the dye acts as a catalyst. For advanced learners, rate law equations can be used to explain the debluing time in relation to concentrations of the three compounds. By measuring the debluing time at varying concentrations of reactants, rate constant $k$ can be calculated. Moreover, repeating the $k$ measurement at a second temperature provides enough information to calculate the of activation energy $E_a$.

We also add thermodynamic discussion to the solution preparation step. The temperature change when water is added to NaOH solution and glucose powder is remarkably noticeable. These are good example to complement a thermodynamic lecture on enthalpy of solution (exothermic for NaOH and endothermic for glucose).

Lastly, we introduce concepts of chemical equilibrium to explain the role of NaOH in this reaction. NaOH is included in the rate law because it reacts with glucose to form a reactive glucoside ion in a rapidly reversible equilibrium. Alternatively, we may use Le Chatelier's principle to explain that NaOH reacts with gluconic acid to form sodium gluconate and hence drives the total reaction to the right.

**3.2 Analytical chemistry**

We have used this experiment to introduce four topics of discussion: concentration calculations, solution preparation, solution transfer, and redox/acid-base indicators. Firstly, users are encouraged to calculate the concentrations of the four chemicals when water is filled into bottles. Secondly, users will need to prepare the four stock solutions by filling water to the marked level on the bottle in the same manner that it is done to a volumetric flask. It is suggested in the booklet that student should (1) fill water close to mark, (2) stir the solution, let the solid dissolved and let the solution equilibrated to the room temperature and (3) complete final volume adjustment using a syringe and mix the solution by closing the cap and inverting the bottle. Thirdly, users learn to transfer solution by using a syringe (avoid air bubbles in the syringe and read the volume at eye level). Finally, concepts of redox and acid-base indicators may be introduced. Indigo carmine is an example of both acid-base and redox indicator. Indigo carmine solution changes from blue to yellow in NaOH solution without adding the sugar.

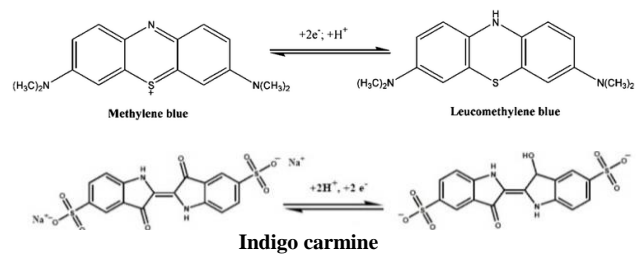

**Indigo carmine**

**3.3 Organic chemistry**

Students are encouraged to explore the relationship between structure and property of involved chemical species. The first obvious example is the conjugation system in MB and leucomethylene blue (LMB). LMB has two separate benzene rings that do not absorb visible light and as a result it is colorless. MB on the other hand, has a long conjugation system of three six-membered rings and shows a deep blue color because of visible light absorption. Another point of discussion is the relationship between the structure of sugars and rate of debluing reaction. Reducing sugars have two anomers (α and β) and the β structure (equatorial OH) oxidizes faster than the α one [5].

It is also beneficial to discuss oxidation numbers of various atoms in the reaction. This helps foster the concepts of redox reaction for organic compounds and complement mainstream view that the process is only functional group transformation. By using the oxidation number argument, students should be able to relate oxidation of glucose to combustion of hydrocarbon or other carbohydrate. (See Biological chemistry section.) The lowest oxidation number for carbon is −4 in $CH_4$ while the highest of +4 is observed in $CO_2$. As nutrients are metabolized in living organisms, the oxidation number of carbon increases along the process until it reaches +4. However, we only observe oxidation of aldehyde (+1) to carboxylic acid (+3) in this reaction.

**3.4 Biological chemistry**

We compare the experiment to animal respiration and combustion. The dye, MB or indigo carmine plays a similar role to hemoglobin (dark red/bright red) or hemocyanin (colorless/blue) in animal blood by carrying oxygen for glucose oxidation. (This is technically not true but acceptable analogy.) Since the dye and oxygen carrying protein facilitate the reaction but not consumed in the total reaction, they are catalyst for oxidation of glucose.

Another line of inquiry is on the combustion process whose chemical equation is the same as cellular respiration.

$$CH_2O + O_2 \rightarrow CO_2 + H_2O$$

Combustion generally occurs at high temperature and is difficult to control. However, the same chemical transformation is also made by living organisms under relatively mild condition, narrow-range of temperature (especially for warm-blooded animals) and controllable energy output. This is an open topic for discussion and there are much more to learn from intricate enzyme catalyzed reactions in living organisms.

When a thin layer of the solution is poured on the petri dish a pattern is formed after it is left for a few minutes. This convection cell is similar to what is observed in boiling liquid [10] and perhaps the discussion can be extended to pattern formation in biology [11].

### 3.5 Scientific method

In the first part of the booklet, we encourage users to formulate hypothesis and propose ways to prove or disprove their ideas. Later on we introduce concept of experimental design in the kinetic investigation. The concentration of components of the solution is varied one at a time to investigate their effect on the debluing reaction. This helps teaching the concepts of independent, dependent and control variables to students. For advance learners, they will be able to plan their own experiments to obtain $E_a$ value. (See Physical chemistry sections.)

### 3.6 Laboratory safety

The NaOH solution is probably the most dangerous chemical in this kit. We stress to students that the safety of the people is of paramount importance. Precautions should be taken before conducting any experiments. Standard procedures involving MSDS and PPE are introduced in the booklet before any experiment-related contents. Eye protection is highly recommended when handling chemicals.

### 4. Preliminary evaluation

Table 1: Average survey scores for ICCH224

| Question | Section I | Section II |
|---|---|---|
| Q1 | 0.313 | 0.667 |
| Q2 | 0.375 | 0.833 |
| Q3 | 0.938 | 0.833 |
| Q4 | 0.313 | 0.333 |
| **Q1-Q4 total** | **1.94** | **2.67** |
| Q5 (satisfaction) | 3.81 | 3.17 |

| List of questions |
|---|
| 1. Do you think the air in the bottle acts as a catalyst? |
| 2. By increasing the concentration of glucose, what will happen to the rate of debluing reaction? Why? |
| 3. If you were to do the blue bottle experiment at a higher temperature do you think that the rate for debluing reaction will increase? |
| 4. What will happen if you reduce the MB concentration? |
| 5. Rate your overall satisfaction of this experiment from 1 (most dissatisfied) to 5 (most satisfied). |

The experiment was given to two sections of ICCH224 Integrated Laboratory Techniques in Chemistry at MUIC. The first and the second sections have 16 and 6 students respectively. During the second week of T1/2014-15, both sections followed Cook's sixteen-question student's worksheet. Only the second section completed experiments from the colorful chemical bottle kit during the third week. Students were asked to complete five multiple-choice questions during the fourth week. The percentages of the students answering the post activity question correctly are listed in Table 1.

Students from section II have higher overall score than section I (2.67 vs. 1.94), but have a slightly lower satisfaction score than section I (3.17 vs. 3.81). Four students in section II took General Chemistry II midterm exam that included two questions on kinetics. The mean scores for the class for the two questions were 0.58±0.34 and 0.23±0.33 while the mean scores for the four students were 0.89 and 0.72 respectively.

### 5. Conclusions

Inspired by earlier work of Campbell and Cook, the colorful chemical bottle experiment kit has been introduced in 2006. The experiment was made more convenient for the general public and quantitative aspect of the reaction can also be explored in detail. Preliminary evaluation at MUIC suggests that the experiment kit is an effective teaching tool for undergraduate students.

### Acknowledgements

We thank two summer students from MUIDS, Tanapon Luechai and Punyawee Lertworawut for their assistance in this project. We appreciate ICCH224 students taking their time to answer our survey and to provide feedbacks. The generous and continued support of NSTDA and its staff is also acknowledged.

### References

[1] Campbell, J., 1963, *J. Chem. Educ.* 40, 578–583.
[2] Engerer, S.C. and Cook, A.G., 1999, *J. Chem. Educ.* 76, 1519–1520.
[3] Shakhashiri, B.Z., 1985, Chemical Demonstrations: A Handbook for Teachers of Chemistry, Vol. 2, Univ of Wisconsin Press.
[4] Chen, P.S., 1970, *J. Chem. Educ.* 47, A335.
[5] Cook, A.G., Tolliver, R.M. and Williams, J.E., 1994, *J. Chem. Educ.* 71, 160–161.
[6] Wellman, W.E., Noble, M.E. and Healy, T., 2003, *J. Chem. Educ.* 80, 537–540.
[7] Adamcikova, L. and Sevcik, P., 1998, *J. Chem. Educ.* 75, 1580.
[8] Jang, N.Y. and Won, K., 2014, *Int. J. Food Sci. Technol.* 49, 650–654.
[9] Schafer, P., van de Linde, S., Lehmann, J., Sauer, M. and Doose, S., 2013, *Anal. Chem.* 85, 3393–3400.
[10] Bees, M., Pons, A., Sørensen, P.G. and Sagues, F., 2001, *J. Chem. Phys.* 114, 1932–1943.
[11] Turing, A.M., 1952, *Philos. Trans. R. Soc. Lond. B. Biol. Sci.* 237, 37–72